\documentclass[12pt,preprint]{aastex}

\usepackage{amsmath}

\slugcomment{Submitted to ApJ}

\shorttitle{Neutrino Heating Between Neutron Stars}
\shortauthors{Salmonson and Wilson}

\begin{document}


\title{Neutrino Annihilation between Binary Neutron Stars}


\author{Jay D. Salmonson}
\author{James R. Wilson}
\affil{Lawrence Livermore National Laboratory, Livermore, CA 94550}



\begin{abstract}

We calculate the neutrino pair annihilation rate into electron pairs
between two neutron stars in a binary system.  We present a closed
formula for the energy deposition rate at any point between the stars,
where each neutrino of a pair derives from each star, and compare this
result to that where all neutrinos derive from a single neutron star.
An approximate generalization of this formula is given to include the
relativistic effects of gravity.  We find that this inter-star
neutrino annihilation is a significant contributor to the energy
deposition between heated neutron star binaries.  In particular, for
two neutron stars near their last stable orbit, inter-star neutrino
annihilation energy deposition is almost equal to that of single star
energy deposition.

%{\bf Note:} Remember to mention that GR is added. We currently assume
%that a neutrino only feels the gravity from its respective NS!  That
%is wrong, we should at least add the star's Schwarzchild metrics (see
%Eqn (8)).  That is still approximate, but more consistant.

\end{abstract}




\section{Introduction}

Neutrino emission from, and interactions above neutron stars has long
been understood to be an important process in supernovae
\citep[e.g.][]{wm93}.  One interaction of particular interest is
neutrino annihilation: $\nu + {\overline{\nu}} \rightarrow e^+ + e^-$
because it converts neutrinos into electrons, which, due to their
vastly larger cross-section, are an important source of energy for
supernovae and possibly gamma-ray bursts.  Because of the strong
gravity of neutron stars, relativistic effects must be included when
calculating neutrino processes.  Rates of the $\nu {\overline{\nu}}
\rightarrow e^+e^-$ interaction above a neutron star were originally
calculated without gravity by \cite{cvb87} and \cite{gdn87}.  In
\cite{sw99} these neutrino annihilation rates were calculated with
gravity and it was found that general relativistic effects
significantly augmented the efficiency; by up to a factor of four for
supernovae and up to a factor of thirty for lone neutron stars at
their critical density.

Another situation where neutrino emission from neutron stars is
important is in neutron star binary (NSB) systems.  Binary neutron
star mergers have long been a possible progenitor for gamma-ray bursts
(GRBs) \cite[e.g.][]{elps89,mr92b}.  In particular, current trends
would suggest that NSBs may be the progenitor of the ``short'' GRBs,
characterized by timescales $\lesssim 1$ second, which are thought to
be distinct from the class of ``long'' GRBs \citep[e.g.][]{fwh99}.  As
such, we are moved to study mechanisms for energy extraction.

Due to the extreme high densities inside the neutron stars, neutrinos,
with their small cross-section, are a necessary component of the
energetics of supernovae and compact binary merger scenarios.  In
particular, only a small proportion ($\sim 1$\%) of the total energy
(which could be as high as the binding energy of a solar mass neutron
star) in neutrinos need be transferred to conventional matter in order
to produce large, observable, energetic events.  Such a mechanism has
been suggested as an energy source for GRBs by
\cite{swm01,wsm97,mr92b,elps89}.

In this paper we study the inter-star neutrino annihilation; where
each neutrino in an annihilation pair derives from each star in the
binary.  This is distinct from the single star neutrino annihilation
discussed in \cite{sw99}.  Both processes will occur in a binary star
system and will sum to give the total energy depostion from neutrino
annihilation.  We will compare and contrast the contributions of these
two processes.  We find that the inter-star component is a very
significant contributor to the total energy deposition.  In Section
\ref{derivationsection} we derive the results without general
relativistic effects.  In Section \ref{generalrelativitysection} we
make some estimates of the augmentation of annihilation due to general
relativity and compare with the relativistic augmentation derived in
\cite{sw99} for the single star case.

\section{Interstar Neutrino Annihilation} \label{derivationsection}

Starting from references in \cite{sw99}, the energy deposition via
neutrino-antineutrino annihilation into electron-positron pairs per
unit time per unit volume is
\begin{equation}
\dot{q} = \frac{D G^2_F }{ 3\pi c^5} {\bf \Theta}(\vec{r}) \iint f_\nu 
f_{\overline{\nu}} (\varepsilon_\nu + 
\varepsilon_{\overline{\nu}}) \varepsilon_\nu^3 d\varepsilon_\nu
\varepsilon_{\overline{\nu}}^3 d\varepsilon_{\overline{\nu}} ~,
\label{qdotdefeqn}
\end{equation}
where $f_\nu$ and $f_{\overline{\nu}}$ are the respective neutrino and
anti-neutrino number densities in phase space, $\varepsilon_\nu$ and
$\varepsilon_{\overline{\nu}}$ are their energies, ${\bf \Theta}
(\vec{r})$ is the integral over incident neutrino angles, $G^2_F =
5.29 \times 10^{-44}$ cm$^2$ MeV$^{-2}$ and
\begin{equation}
D = 1 \pm 4 \sin^2\theta_W + 8 \sin^4 \theta_W ~,
\label{E:D}
\end{equation}
where the $+$ sign is for $\nu_e{\overline{\nu}}_e$ pairs, while $-$
is for $\nu_\mu\overline{\nu}_\mu$ and $\nu_\tau{\overline{\nu}}_\tau$
pairs.  We assume black-body-like neutrino emission; a good assumption
for thermal neutrinos diffusing out of a hot neutron star
\citep{wmm96}.  Neglecting redshift effects on the temperature for the
moment, eqn.~(\ref{qdotdefeqn}) reduces to
\begin{equation}
\dot{q} = 7.97 \times 10^{21} D~ \biggl( \frac{T}{1 \text{MeV}} \biggr)^9
{\bf \Theta}(\vec{r})~ \text{ergs cm}^{-3}~ \text{sec}^{-1}
\label{qdoteqn}
\end{equation}
where $D = 1.23$ for $\nu_e{\overline{\nu}}_e$ pairs and otherwise
$D = 0.814$.  The integration over incident neutrino angles is
\begin{equation}
{\bf \Theta}(\vec{r}) \equiv \iint (1 - {\boldsymbol \Omega}_\nu \cdot 
{\boldsymbol \Omega}_{\overline \nu})^2 d\Omega_\nu d\Omega_{\overline \nu} ~.
\label{E:Theta}
\end{equation}

We define the neutrino trajectories, in $(x,y,z)$, from star 1 (on the
left) as
\begin{equation}
{\boldsymbol \Omega}_1 = (\cos \theta'_1,~ \sin \theta'_1
\cos \phi'_1,~ \sin \theta'_1 \sin \phi'_1) 
\end{equation}
and for star 2 (on the right) as (note that $x \rightarrow -x$ do to
reflection symmetry)
\begin{equation}
{\boldsymbol \Omega}_2 = (- \cos \theta'_2,~ \sin \theta'_2
\cos \phi'_2,~ \sin \theta'_2 \sin \phi'_2) 
\end{equation}
where $(\theta'_1,\phi'_1)$ and $(\theta'_2,\phi'_2)$ are the
direction angles for the neutrino trajectories in spherical
coordinates from stars 1 and 2 respectively, so that $(\theta'_i = 0,
\phi'_i =0)$ points to the center of star $i = 1,2$. Since we wish to
evaluate the integral of eqn. (\ref{E:Theta}) at any point in space
around the stars, we rotate the vector ${\boldsymbol \Omega}_1$ by an
angle $\Theta_1$
\begin{equation}
{\boldsymbol \Omega}_1 = (\cos \theta'_1 \cos \Theta_1 - \sin \theta'_1
\cos \phi'_1 \sin \Theta_1,~ \sin \theta'_1 \cos \phi'_1 \cos \Theta_1 +
\cos \theta'_1 \sin \Theta_1,~ \sin \theta'_1 \sin \phi'_1) ~.
\end{equation}
A rotation of ${\boldsymbol \Omega}_2$ by $\Theta_2$ gives (note that
${\boldsymbol \Omega}_2$ will have the same form as ${\boldsymbol
\Omega}_1$ except $x \rightarrow -x$)
\begin{equation}
{\boldsymbol \Omega}_2 = (- \cos \theta'_2 \cos \Theta_2 + \sin
\theta'_2 \cos \phi'_2 \sin \Theta_2,~ \sin \theta'_2 \cos \phi'_2 \cos
\Theta_2 + \cos \theta'_2 \sin \Theta_2,~ \sin \theta'_2 \sin \phi'_2) ~.
\end{equation}

\begin{figure}
\plotone{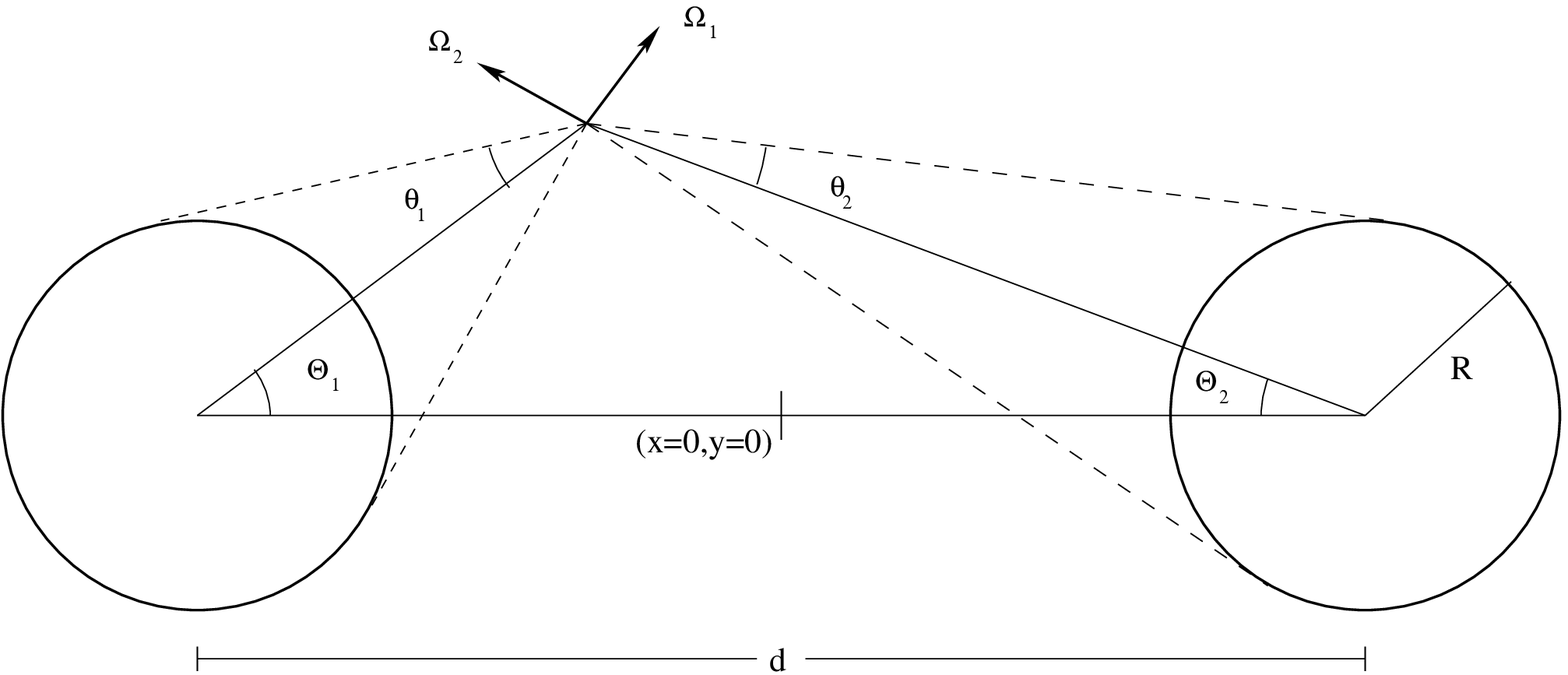}
\caption{Diagram representing the geometry and defining quantities
used in Eqns (\ref{bigeqn}, \ref{smallthetadef}, \ref{bigThetadef}).
For clarity, this illustration shows only the special case where the
neutrino trajectories are coplanar (i.e.~$\phi_1 = \phi_2 =
0$). \label{stargeom}}
\end{figure}

\begin{figure}
\plotone{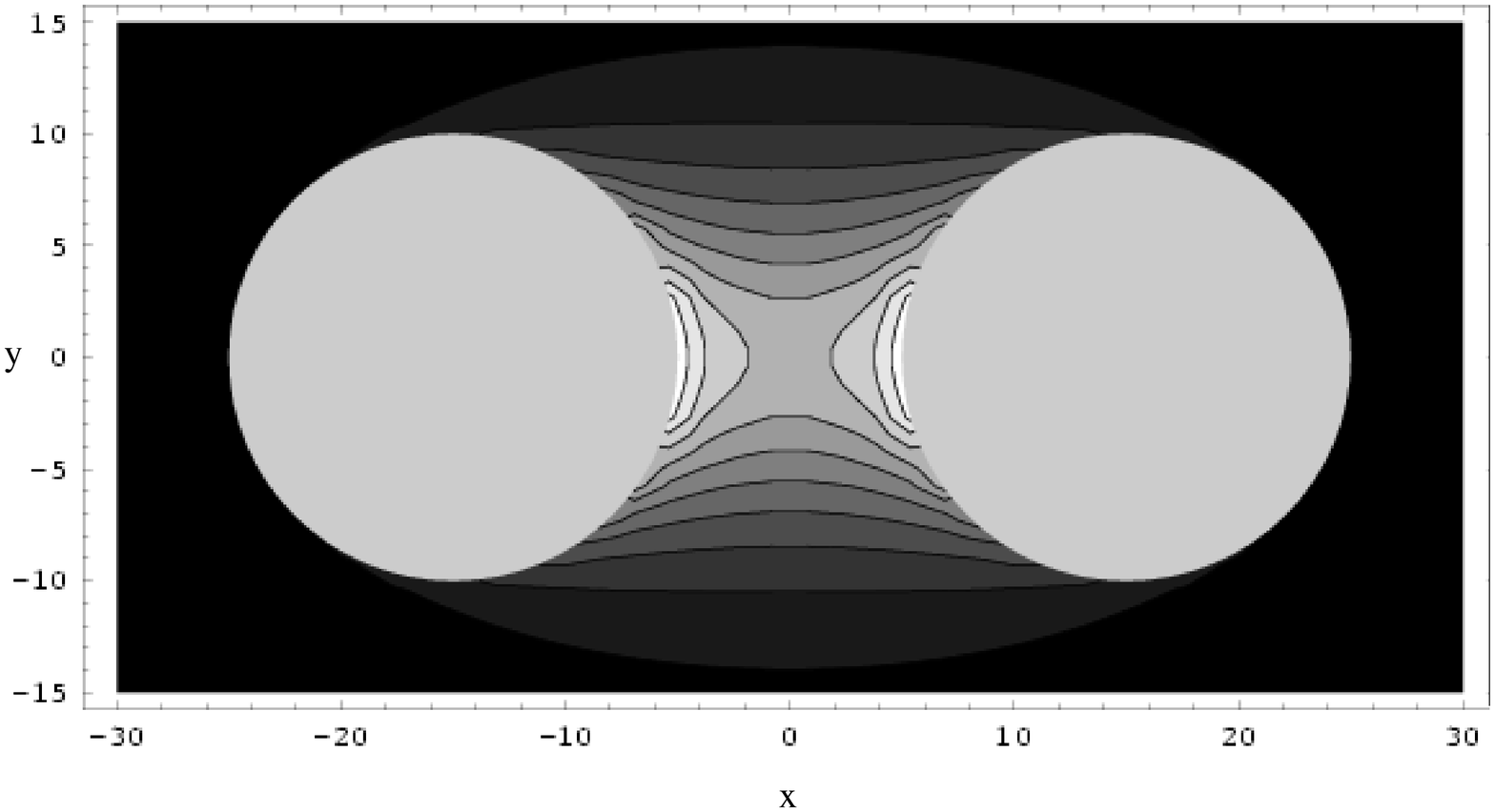}
\caption{A contour plot of a slice through the stars of the angular
dependence of the inter-star neutrino energy deposition density, ${\bf
\Theta}_{inter}$, given by eqn.~(\ref{bigeqn}) with Eqns
(\ref{smallthetadef}, \ref{bigThetadef}), with stellar radii 10 km and
separation 30 km. There are ten 1.1 increment contours starting from
black, (0,1.1), and increasing to white.  The neutron stars are
represented by grey disks.  This figure is meant to illustrate the
function ${\bf \Theta}_{inter}$ and thus does not include eclipsing of
neutrino paths by the neutron stars which would attenuate this
function around the far sides of the neutron stars.  Since the effect
is already small in these regions, this modification is negligible.
\label{contour}}
\end{figure}

Thus, noting $(1 - {\boldsymbol \Omega}_1 \cdot {\boldsymbol
\Omega}_2)^2 = 1 - 2 {\boldsymbol \Omega}_1 \cdot {\boldsymbol
\Omega}_2 + ({\boldsymbol \Omega}_1 \cdot {\boldsymbol \Omega}_2)^2$
and $d \Omega_i = \sin \theta'_i d \theta'_i d \phi'_i$ where $i =
1,2$, eqn.~(\ref{E:Theta}) can be integrated and becomes

\begin{equation}
\begin{split}
&{\bf \Theta}_{\text{inter}}(\theta_1, \theta_2, \Theta_1, \Theta_2) =
 (\pi^2 (516 + 8 \cos[\theta_1] + 4 \cos[2 \theta_1] + 4 \cos[\theta_1
 - 2 \theta_2] + 8 \cos[\theta_1 - \theta_2] + 2 \cos[2 (\theta_1 -
 \theta_2)] \\ & + 4 \cos[2 \theta_1 - \theta_2] + 8 \cos[\theta_2] +
 4 \cos[2 \theta_2] + 8 \cos[\theta_1 + \theta_2] + 2 \cos[2 (\theta_1
 + \theta_2)] + 4 \cos[2 \theta_1 + \theta_2] \\ & + 4 \cos[\theta_1 +
 2 \theta_2] + 12 \cos[\theta_1 - 2 \Theta_1 - \theta_2 - 2 \Theta_2]
 + 6 \cos[2 \theta_1 - 2 \Theta_1 - \theta_2 - 2 \Theta_2] \\ & + 12
 \cos[\theta_1 - 2 \Theta_1 + \theta_2 - 2 \Theta_2] + 6 \cos[2
 \theta_1 - 2 \Theta_1 + \theta_2 - 2 \Theta_2] + 96 \cos[\theta_1 -
 \Theta_1 - \Theta_2] \\ & + 6 \cos[2 (\theta_1 - \Theta_1 -
 \Theta_2)] + 48 \cos[\theta_1 - \Theta_1 - \theta_2 - \Theta_2] + 3
 \cos[2 (\theta_1 - \Theta_1 - \theta_2 - \Theta_2)] \\ & + 48
 \cos[\theta_1 - \Theta_1 + \theta_2 - \Theta_2] + 3 \cos[2 (\theta_1
 - \Theta_1 + \theta_2 - \Theta_2)] + 192 \cos[\Theta_1 + \Theta_2] +
 12 \cos[2 (\Theta_1 + \Theta_2)] \\ & + 96 \cos[\theta_1 + \Theta_1 +
 \Theta_2] + 6 \cos[2 (\theta_1 + \Theta_1 + \Theta_2)] + 96
 \cos[\Theta_1 - \theta_2 + \Theta_2] + 6 \cos[2 (\Theta_1 - \theta_2
 + \Theta_2)] \\ &+ 48 \cos[\theta_1 + \Theta_1 - \theta_2 + \Theta_2]
 + 3 \cos[2 (\theta_1 + \Theta_1 - \theta_2 + \Theta_2)] + 96
 \cos[\Theta_1 + \theta_2 + \Theta_2] \\ & + 6 \cos[2 (\Theta_1 +
 \theta_2 + \Theta_2)] + 48 \cos[\theta_1 + \Theta_1 + \theta_2 +
 \Theta_2] + 3 \cos[2 (\theta_1 + \Theta_1 + \theta_2 + \Theta_2)] \\
 & + 12 \cos[2 \Theta_1 - \theta_2 + 2 \Theta_2] + 12 \cos[\theta_1 +
 2 \Theta_1 - \theta_2 + 2 \Theta_2] + 6 \cos[2 \theta_1 + 2 \Theta_1
 - \theta_2 + 2 \Theta_2] \\ &+ 12 \cos[2 \Theta_1 + \theta_2 + 2
 \Theta_2] + 12 \cos[\theta_1 + 2 \Theta_1 + \theta_2 + 2 \Theta_2] +
 6 \cos[2 \theta_1 + 2 \Theta_1 + \theta_2 + 2 \Theta_2] \\ &+ 12
 \cos[\theta_1 - 2 (\Theta_1 + \Theta_2)] + 12 \cos[\theta_1 + 2
 (\Theta_1 + \Theta_2)] + 6 \cos[\theta_1 - 2 (\Theta_1 - \theta_2 +
 \Theta_2)] \\ &+ 6 \cos[\theta_1 + 2 (\Theta_1 - \theta_2 +
 \Theta_2)] + 6 \cos[\theta_1 - 2 (\Theta_1 + \theta_2 + \Theta_2)] \\
 &+ 6 \cos[\theta_1 + 2 (\Theta_1 + \theta_2 + \Theta_2)])
 \sin^2[\theta_1/2] \sin^2[\theta_2/2])/24 ~.
\end{split}
\label{bigeqn}
\end{equation}
An illustration of the geometry is shown in Figure \ref{stargeom}.
Specifically, $\theta_i$ is the apparent radial angular size of star
$i$ and $\Theta_i$ is the angle star $i$ is observed at with respect
to the axis connecting the star centers. It is useful to convert
eqn.~(\ref{bigeqn}) to cartesian coordinates (${\bf \Theta}(\theta_1,
\theta_2, \Theta_1, \Theta_2) \rightarrow {\bf \Theta}(x,y)$) via the
following substitutions.  If two stars of mass $M$ and radius $R$ are
separated by distance $d$, so the half-separation is $d_2 \equiv d/2$
and

\begin{equation}
r_{1,2} \equiv \sqrt{(d_2 \pm x)^2 + y^2} ~, 
\label{defr}
\end{equation}
then

\begin{equation}
\theta_{1,2} = \arcsin \biggl( \frac{R}{r_{1,2}} \biggr)
%\theta_\pm = \arcsin \Biggl( \frac{ r_{\text{s}}}{r_{\pm}}
%	\sqrt{\frac{ 1 - 2 M/r_{\pm}}{ 1 - 2 M/r_{\text{s}}}} \Biggr)
\label{smallthetadef}
\end{equation}

\begin{equation}
\Theta_{1,2} = \arctan \biggl( \frac{y}{d_2 \pm x} \biggr)
\label{bigThetadef}
\end{equation}
where the $+$ and $-$ signs correspond to stars to the left (star 1)
and right (star 2) of the axis of symmetry respectively.  Figure
\ref{contour} shows a contour plot of this integration over inter-star
incident neutrino angles given by eqns.~(\ref{bigeqn},
\ref{smallthetadef}, \ref{bigThetadef}).  As expected, $\nu
\overline{\nu}$ annihilation takes place primarily between the stars,
where the cross-section and flux are highest.

\begin{figure}
\plotone{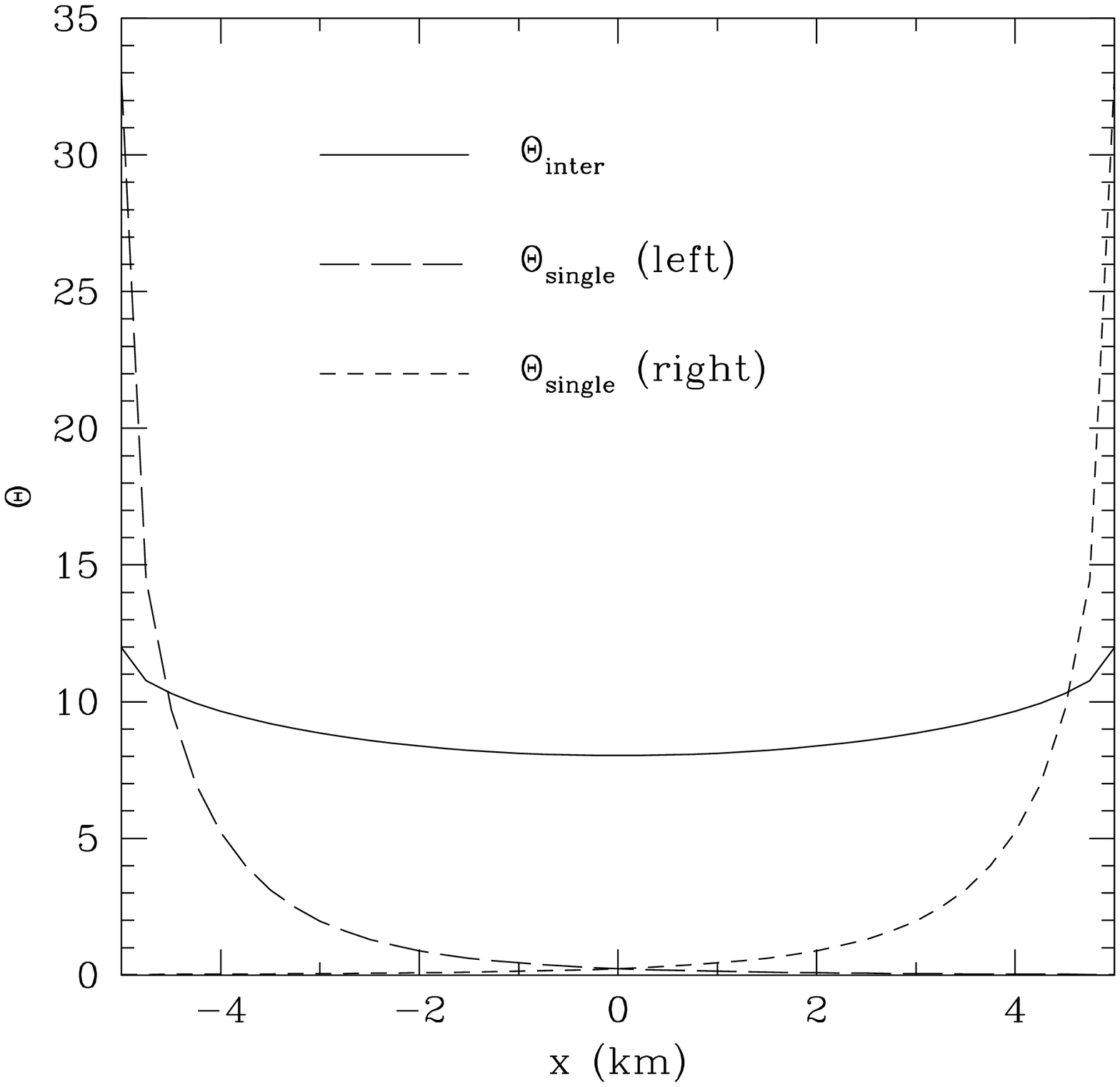}
\caption{A plot of the region between the two neutron stars along the
line connecting the two star centers.  The left (right) edge of the
plot is the surface of the left (right) star.  The solid line is the
inter-star deposition ${\bf \Theta}_{\text{inter}}$ given by Eqn
(\ref{bigeqn}, \ref{smallthetadef}, \ref{bigThetadef}) and shown also
in Figure \ref{contour}.  The dashed lines are the single star angular
deposition functions ${\bf \Theta}_{\text{single}}$
(Eqn. \ref{E:Thetaresult}, \ref{E:defx}, Newtonian; $M=0$) for each
star of radius 10 km and separation 30 km. Note that ${\bf
\Theta}_{\text{single}}$ decays rapidly with distance from the surface
of each neutron star while ${\bf \Theta}_{\text{inter}}$ is more
gradually varying between the stars.
\label{betweenstars}}
\end{figure}

For comparison, the result for a single neutron star is \citep{sw99}
\begin{equation}
{\bf \Theta}(r)_{\text{single}} = \frac{2 \pi^2}{3} (1 - x)^4 (x^2 + 4 x + 5) 
\label{E:Thetaresult}
\end{equation}
where $r$ is the distance from the neutron star center and, for a star of mass $M$ and radius $R$, we define
\begin{equation}
x \equiv \sqrt{ 1 - \biggl(\frac{R}{r}\biggr)^2 
\frac{1 - \frac{2M }{ r} }{ 1 - \frac{2M}{ R}} } ~.
\label{E:defx}
\end{equation}
This equation (\ref{E:defx}) takes into account the effects of general
relativity in Schwarzchild coordinates. In particular, the
gravitational bending of neutrino trajectories (see Section
\ref{generalrelativitysection}).  For the Newtonian result, take $M =
0$.  In Figure \ref{betweenstars} is shown a comparison between ${\bf
\Theta}_{\text{inter}}$ (eqn. \ref{bigeqn}) and ${\bf
\Theta}_{\text{single}}$ (eqn. \ref{E:Thetaresult}) along the axis
between the stars.  We see that annihilation of $\nu \overline{\nu}$
from a single star is dominant near the stellar surfaces, while
inter-star annihilation is important between the stars.

\begin{figure}
\plotone{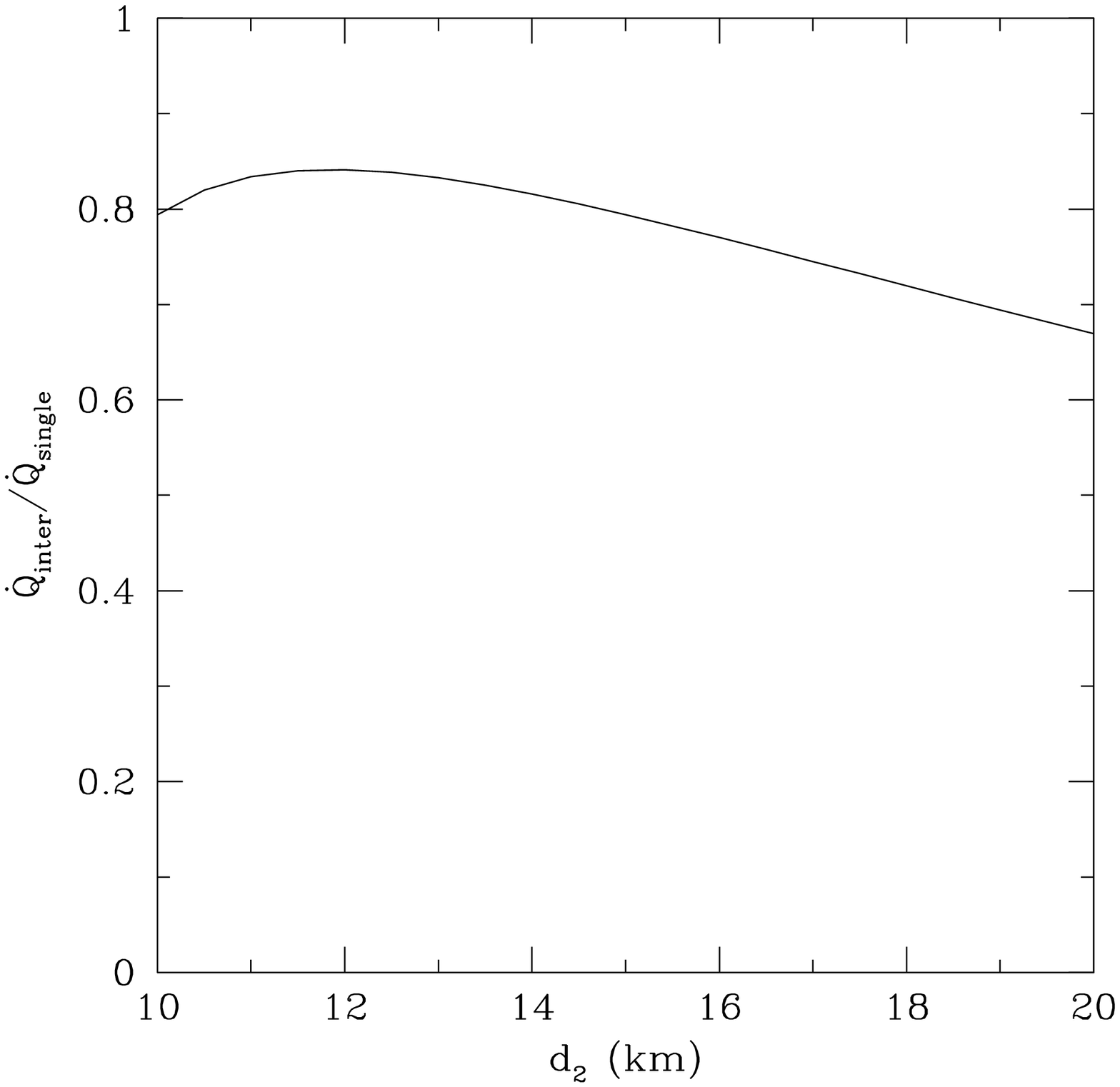}
\caption{Ratio of integrals over space of inter-star energy angle
factor (Eqns \ref{bigeqn}, \ref{smallthetadef}, \ref{bigThetadef}) to
single star angle factor (Eqns \ref{E:Thetaresult}, \ref{E:defx},
Newtonian; $M=0$).  This is evaluated for star radii 10 km over a
range of half-distances, $d_2$, between the star centers. The
turn-down for small separations ($d_2 < 12$ km) is due to the paucity
of volume between the stars. \label{ratiorange}}
\end{figure}

One can compare the rate of inter-star deposition to single star
deposition.  A key conclusion of this paper is that the inter-star
annihilation is a significant component of the total energy
deposition.  In Figure \ref{ratiorange} we see a ratio of total energy
deposited via inter-star annihilation versus that for a single star
over a range of stellar separations.  The total energy deposited is an
integral of eqn.~(\ref{qdoteqn}) over the interstellar volume $V$
\begin{equation}
\dot{Q}_{\text{Newt}} \equiv \int \dot{q} ~dV \propto \int {\bf \Theta} ~ dV ~.
\label{Qdoteqn}
\end{equation}
We see that in this Newtonian limit
$\dot{Q}_{\text{inter}}/\dot{Q}_{\text{single}} \sim 80$\% for close
separations (i.e.~$d_2 < 2 R$) where $\dot{Q}_{\text{inter}} \propto \int
{\bf \Theta}_{\text{inter}}~dV$ using eqn.~(\ref{bigeqn}) and
$\dot{Q}_{\text{single}} \propto \int {\bf \Theta}_{\text{single}}~dV$
using eqn.~(\ref{E:Thetaresult}).

\section{ General Relativity } \label{generalrelativitysection}

% define isotropic r:
\def\olr{\overline{r}}
\def\olR{\overline{R}}

Analytical calculation of the general relativistic effects on
inter-star neutrino annihilation is extremely complex.  Instead, we
make some crude approximations to gain some insight into the effect of
gravity.  We neglect possible rotation effects.  It is most convenient
and intuitive to use isotropic coordinates
\begin{equation}
ds^2 = - \alpha^2 dt^2 + \phi^4 (d\olr^2 + \olr^4 d\Omega^2)
\label{isotropicmetric}
\end{equation}
where, for one star \citep{mtw73}
\begin{equation}
\begin{split}
\alpha &= \frac{1 - \frac{M}{2 \olr}}{1 + \frac{M}{2 \olr}}  \\
\phi   &= 1 + \frac{M}{2 \olr} 
\end{split}
\end{equation}
and the isotropic distance, $\olr$, to the center of the star
is related to the Schwarzchild distance $r$ by
\begin{equation}
r = \olr \biggl(1 + \frac{M}{2 \olr} \biggr)^2 = \olr \phi(\olr)^2 ~.
\end{equation}

In the two star environment we define
\begin{equation}
\begin{split}
\alpha(\olr_+,\olr_-) &\equiv 1 - \frac{M}{\olr_+} - \frac{M}{\olr_-} \\
\phi(\olr_+,\olr_-)   &\equiv 1 + \frac{M}{2 \olr_+} + \frac{M}{2 \olr_-} 
\end{split}
\label{isotropic2starmetric}
\end{equation}
where $r_+$ ($r_-$) indicates the distance to the center of the star
on the left (right).  This choice of metric is found to be an adequate
approximation, to a few percent, of the real metric as numerically
calculated by \cite{wmm96}.  The value of these metric components at
the stellar surface is not at constant potential (because we neglect
tidal distortions which equalize the surface potential), thus we
define the potentials at the surface to be
\begin{equation}
\alpha_s \equiv \alpha(\olR,d_s) ~, ~ \phi_s \equiv \phi(\olR,d_s)
\end{equation}
with
\begin{equation}
d_s \equiv \frac{\olR}{r_\mp} \sqrt{ \biggl(d_2 \bigl( 2 \frac{r_\mp}{\olR} - 1 \bigr) \pm x \biggr)^2 + y^2}
\end{equation}
using eqn.~(\ref{defr}) and where one chooses the upper sign if $x>0$
and vice versa.  This corresponds to the potential at the surface of
the star at the point for which the line to point ($x,y$) is
perpendicular to the surface.  From here on we simplify the notation
by suppressing the metric arguments (e.g. $\alpha \equiv
\alpha(\olr_+,\olr_-)$).

For the single star case, in these coordinates, eqn.~(\ref{E:defx})
becomes
\begin{equation}
x = \sqrt{ 1 - \biggl(\frac{\olR}{\olr}\biggr)^2
\biggl(\frac{\phi_s}{\phi}\biggr)^3
\biggl(\frac{\alpha}{\alpha_s}\biggr)^2}
%\biggl(\frac{\phi(\olR)}{\phi(\olr)}\biggr)^3
%\biggl(\frac{\alpha(\olr)}{\alpha(\olR)}\biggr)^2}
%x = \sqrt{ 1 - \biggl(\frac{\olR}{\olr}\biggr)^2 
%\biggl(\frac{ 1 + \frac{M}{2 \olR}}{1 + \frac{M }{2 \olr} }\biggr)^6
%\biggl(\frac{1 - \frac{M }{2 \olr }}{ 1 - \frac{M}{2 \olR}}\biggr)^2 } 
\label{E:defxgr}
\end{equation}
where $\olR$ is the isotropic stellar radius. For compact neutron
stars near collapse, we will take $\olR = 7.5$ km \citep{wmm96}.

For the inter-star case, the apparent radial angular size, $\theta$,
of a star will be augmented due to the gravitational bending of
neutrino geodesics by the mass of the star.  We include this effect by
modifying eqn.~(\ref{smallthetadef}) by identifying $R \rightarrow
\olR \phi_s^{3/2}/ \alpha_s$ and $r \rightarrow \olr \phi^{3/2}/\alpha$
from eqn.~(\ref{E:defxgr}).  These substitutions take into account the
``lensing'' effect due to the gravitational bending of the neutrino geodesics
\citep{sw99} and give
\begin{equation}
\theta_\pm \equiv \arcsin \Biggl( \frac{\olR}{\olr_{\pm}}
	\biggl(\frac{\phi_s}{\phi}\biggr)^{3/2}
	\biggl(\frac{\alpha}{\alpha_s}\biggr)
%	\biggl(\frac{\phi(\olR,d)}{\phi(\olr_+,\olr_-)}\biggr)^{3/2}
%	\biggl(\frac{\alpha(\olr_+,\olr_-)}{\alpha(\olR,d)}\biggr)
%	\biggl( \frac{ 1 + \frac{M}{ 2 \olR}}{ 1 + \frac{M}{ 2\olr_{\pm}} 
%	} \biggr)^3
%	\biggl(	\frac{ 1 - \frac{ M }{ 2\olr_{\pm}} }{ 1 -  \frac{M}{ 2 \olR}}
%	 \biggr) 
\Biggr) ~.
\label{smallthetadefgr}
\end{equation}

In addition, there will be a redshift (or blueshift) of the neutrino
energies.  Specifically, assuming the neutrinos emerge from the
neutrinosphere of each star with surface temperature $T_s$, then the
temperature at a position $(x,y)$, using eqn.~(\ref{defr}), is
\begin{equation}
%T(\olr_+,\olr_-) = \frac{\alpha(\olR,d)}{\alpha(\olr_+,\olr_-)} T(\olR,d) ~.
T(\olr_+,\olr_-) = \frac{\alpha_s}{\alpha} T_s ~.
\end{equation}
Then the energy deposited at this point is (eqn.~\ref{qdoteqn})
\begin{equation}
\dot{q} = 7.97 \times 10^{21} D~\biggl( \frac{T(\vec{r})}{1
\text{MeV}} \biggr)^9 {\bf \Theta}(\vec{r})~ \text{ergs cm}^{-3}~
\text{sec}^{-1} \propto \biggl( \frac{\alpha_s}{\alpha}
T_s \biggr)^9 {\bf \Theta}(\vec{r}) ~.
%\text{sec}^{-1} \propto \biggl( \frac{\alpha(\olR,d)}{\alpha(\olr_+,\olr_-)}
%T(\olR,d) \biggr)^9 {\bf \Theta}(\vec{r}) ~.
\end{equation}

\begin{figure}
\plotone{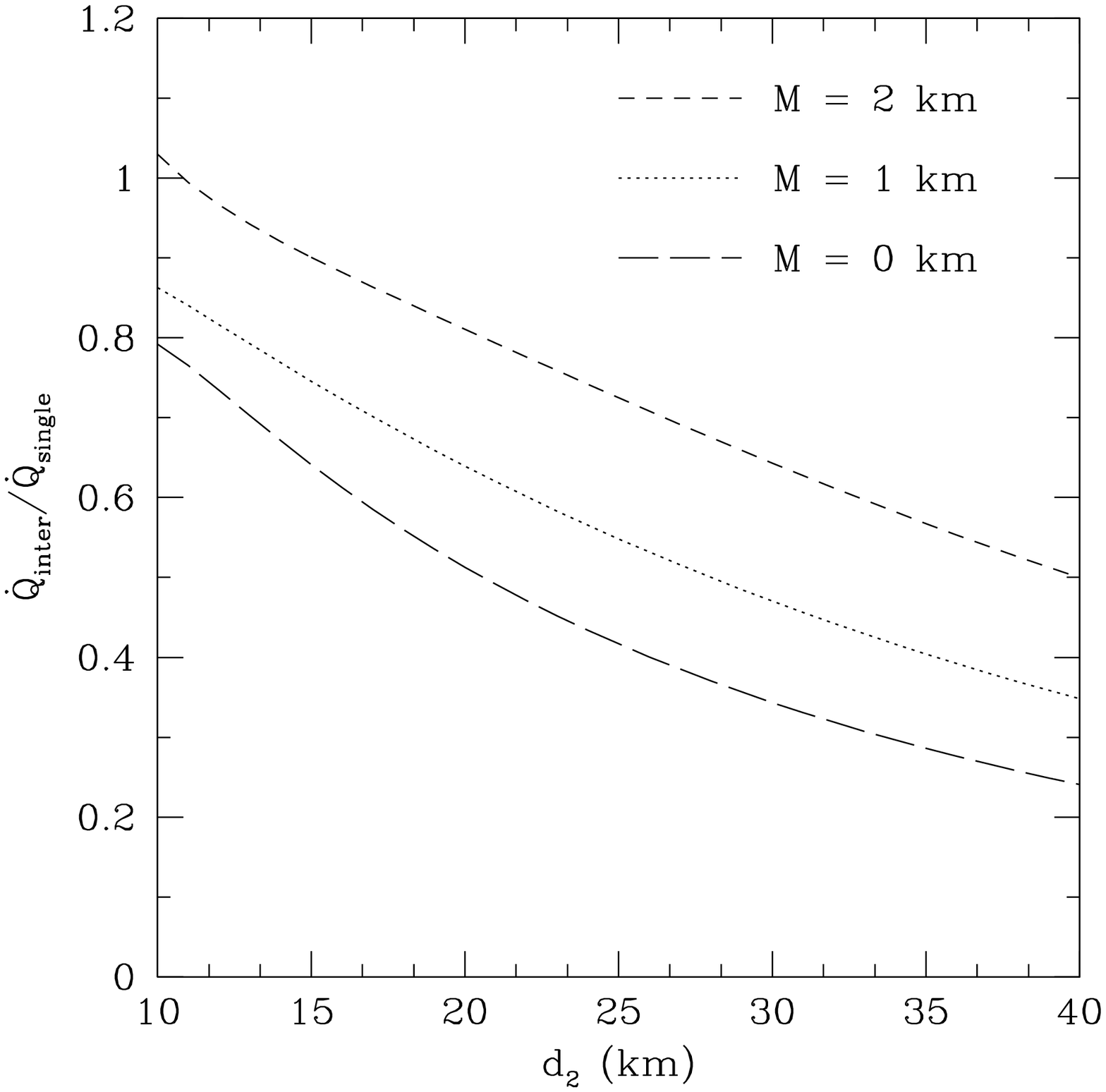}
\caption{A ratio of the entire energy deposition due to inter-star
neutrino annihilations, $\dot{Q}_{\text{inter}}$, to that from single
star annihilations, $\dot{Q}_{\text{single}}$, is shown for isotropic
stellar radii $\olR = 7.5$ km for a range of stellar half-separations,
$d_2$, and for several stellar masses $M$.  $M = 0$ corresponds to the
Newtonian, zero-gravity case.  Note that for $M = 2~\text{km} \approx
1.4 M_\odot$ and $d_2 = 10$ km inter-star energy deposition,
$\dot{Q}_{\text{inter}}$, equals single star energy deposition
$\dot{Q}_{\text{single}}$.
\label{edepgr}}
\end{figure}

\begin{figure}
\plotone{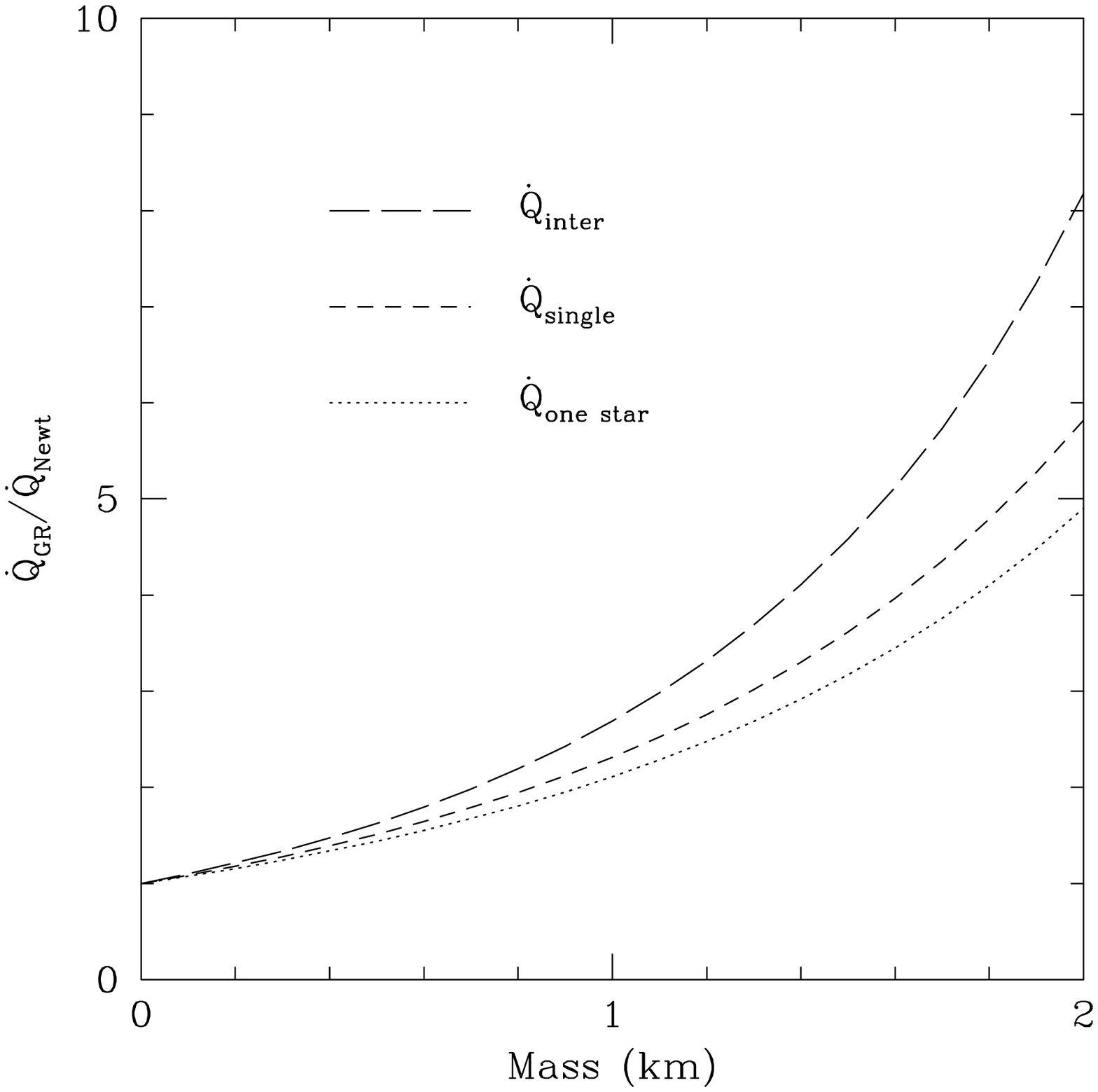}
\caption{A comparison of the ratio
$\dot{Q}_{\text{GR}}/\dot{Q}_{\text{Newt}}$ of general relativistic
(eqn.~\ref{dotQgr}) to Newtonian (eqn.~\ref{Qdoteqn}) energy
deposition.  This ratio is shown for inter-star
$\dot{Q}_{\text{inter}}$ and single star $\dot{Q}_{\text{single}}$
energy depostions for a range of stellar masses and with isotropic
stellar radius $\olR = 7.5$ km and half-separation $d_2 = 15$ km. As
mass increases, both components are augmented by relativistic effects.
For reference, the energy deposition is shown for an isolated neutron
star, $\dot{Q}_{\text{one star}}$, i.e.~the same as
$\dot{Q}_{\text{single}}$, but with the companion star mass set to
zero.  The depositions in a binary system ($\dot{Q}_{\text{inter}}$,
$\dot{Q}_{\text{single}}$) are augmented over $\dot{Q}_{\text{one
star}}$ because of the deeper, two-star potential well.
\label{edepgrm}}
\end{figure}

Following \citet{sw99} we define the total energy deposition
\begin{equation}
\dot{Q}_{\text{GR}} \equiv \int \dot{q} dV = 7.97 \times 10^{21} D~ \biggl(
%\frac{\alpha(\olR,d) T(\olR,d)}{1 \text{MeV}} \biggr)^9 \int
%\frac{{\bf \Theta}(\vec{r})}{\alpha(\olr_+,\olr_-)^9}
%\phi(\olr_+,\olr_-)^6 dx^3 ~ \text{ergs sec}^{-1}
\frac{\alpha_s T_s}{1 \text{MeV}} \biggr)^9 \int
\frac{{\bf \Theta}(\vec{r})}{\alpha^9}
\phi^6 dx^3 ~ \text{ergs sec}^{-1}
\label{dotQgr}
\end{equation}
where the metric is given by eqn.~(\ref{isotropic2starmetric}) and the
inter-star angle factor is calculated using eqns.~(\ref{bigeqn},
\ref{bigThetadef}, \ref{smallthetadefgr}) and the single star angle
factor is given by eqns.~(\ref{E:Thetaresult}, \ref{E:defxgr}).  A
comparison of these two rates for several stellar masses and a range
of separations is shown in Figure \ref{edepgr}.  One can see that as
the stars approach their last stable orbit, $d_2 \sim 10 - 12$ km
\citep{mw00}, the inter-star contribution to neutrino energy
deposition, $\dot{Q}_{\text{inter}}$ becomes comparable to the single
star contribution $\dot{Q}_{\text{single}}$.  Thus the inter-star
energy deposition is significant for hot neutron stars in close
binaries.  In Figure \ref{edepgrm} is shown the augmentation of both
inter-star, $\dot{Q}_{\text{inter}}$, and single star,
$\dot{Q}_{\text{single}}$, energy depositions for a typical
half-separation, $d_2 = 15$ km, and a range of masses.  These are
compared with the energy deposition for an isolated neutron star,
$\dot{Q}_{\text{one star}}$, \citep{sw99} here presented in isotropic
coordinates.

\section{ Discussion  }

Herein we have calculated the inter-star neutrino annihilation rate
for close neutron stars of equal mass in a binary system.  A
generalization of interest would be the consideration of systems with
stars of unequal mass.  While a detailed calculation of such is
difficult and beyond the capability of the analytical results
presented here, we can argue that, at least for small departures from
mass equality, the inter-star annihilation rate will not change much.
In the stellar compression effect discussed by \citep{mw00}, the
larger of the two unequal mass stars may be thought to receive larger
compression due to its excess mass.  From the results given in tables
III and IV of \citet{mw00} is found that the compression for a given
four-velocity is quadratic in mass.  However, the smaller mass star
will move faster, varying inversely with mass, and compression is
shown to scale quadratically with four-velocity \citep{mw00}.
Therefore we expect that there will be a balancing effect keeping the
stars at approximately equal temperatures and thus we expect the
energy deposition of eqn.~(\ref{qdoteqn}), which scales like $\propto
T^9$, to be roughly constant for stars differing in mass by 5 \%,
which is a reasonable range for observed neutron star binaries.

In conclusion, we find that inter-star neutrino annihilation is a
significant component of the total neutrino annihilation in a neutron
star binary system and will thus be a major contributor to the
available energy of coallescing binary systems.  We provide formulas
for calculating this energy deposition.  Future work will use 3-D
relativistic hydrodynamic calculations to take this effect into
account and thus improve the calculations of \cite{swm01}.  In
particular the inter-star annihilation will lead to non-isotropic
outflow.

This work was performed under the auspices of the U.S. Department of
Energy by University of California Lawrence Livermore National
Laboratory under contract W-7405-ENG-48.\\

\clearpage

\end{document}